\documentclass{emulateapj}

\lefthead{A. van der Wel \& R.P. van der Marel}
\righthead{Spatially Resolved Kinematics at $z=1$}
\slugcomment{The Astrophysical Journal, Accepted}

\begin{document}

\title{Spatially Resolved Stellar Kinematics of Field Early-Type
Galaxies at $z=1$: Evolution of the Rotation Rate \altaffilmark{1}}

\author{Arjen van der Wel\altaffilmark{2} \& Roeland P.~van der
Marel\altaffilmark{3}}

\altaffiltext{1}{Based on observations collected at the European
  Southern Observatory, Chile (169.A-0458), and on observations with
  the \textit{Hubble Space Telescope}, obtained at the Space Telescope
  Science Institute, which is operated by AURA, Inc., under NASA
  contract NAS 5-26555.}

\altaffiltext{2}{Department of Physics and Astronomy,
Johns Hopkins University, 3400 North Charles Street, Baltimore, MD
21218; e-mail: wel@pha.jhu.edu}

\altaffiltext{3}{Space Telescope Science Institute, 3700 San Martin
  Drive, Baltimore 21218}

\newcommand{\lta}{\lesssim}
\newcommand{\gta}{\gtrsim}
\newcommand{\kms}{\>{\rm km}\,{\rm s}^{-1}}

\begin{abstract}
  We use the spatial information of our previously published VLT/FORS2
  absorption line spectroscopy to measure mean stellar velocity and
  velocity dispersion profiles of 25 field early-type galaxies at a
  median redshift $z=0.97$ (full range $0.6<z<1.2$). This provides the
  first detailed study of early-type galaxy rotation at these
  redshifts. From surface brightness profiles from \textit{HST}
  imaging we calculate two-integral oblate axisymmetric Jeans equation
  models for the observed kinematics. Fits to the data yield for each
  galaxy the degree of rotational support and the mass-to-light ratio
  $M/L_{\rm Jeans}$. S0 and Sa galaxies are generally rotationally
  supported, whereas elliptical galaxies rotate less rapidly or not at
  all. Down to $M_B=-19.5$ (corrected for luminosity
  evolution), we find no evidence for evolution in the fraction of
  rotating early-type (E+S0) galaxies between $z\sim 1$ ($63\%\pm
  11\%$) and the present ($61\%\pm 5\%$). We interpret this as
  evidence for little or no change in the field S0 fraction with
  redshift.  We compare $M/L_{\rm{Jeans}}$ with $M/L_{\rm{vir}}$
  inferred from the virial theorem and globally averaged quantities,
  and assuming homologous evolution. There is good agreement for
  nonrotating (mostly E) galaxies. However, for rotationally
  supported galaxies (mostly S0) $M/L_{\rm{Jeans}}$ is on average
  $\sim 40\%$ higher than $M/L_{\rm{vir}}$.  We discuss possible
  explanations and the implications for the evolution of $M/L$ between
  $z=1$ and the present and its dependence on mass.
\end{abstract}


\section{INTRODUCTION}
\label{intro}

Detailed studies of nearby early-type galaxies provide insight into
their stellar populations, dark matter content and kinematic and
spatial structure. In particular, resolved kinematic data reveal the
underlying gravitational potential \citep[e.g.,][]{vandermarel91,
cappellari06}, the presence of super-massive black holes
\citep[e.g.,][]{richstone90, verolme02} and dark halos
\citep[e.g.,][]{kronawitter00}, and the relative contributions of
pressure and rotation to its orbital energy \citep[e.g.,][]{binney78,
davies83, emsellem07}.

Recently, the first, probing steps have been made toward spatially
resolving the stellar motions in more distant early-type galaxies
using absorption-line spectroscopy. The past few years have seen
studies aimed at constraining the dark halos of lensing galaxies
\citep{koopmans06}, dynamically distinguishing cluster E and S0
galaxies \citep{moran07b}, constraining the evolution in rotation rate
\citep{vandermarel07a}, and constraining the evolution of the
mass-to-light ratio ($M/L$) using the $M/L$ versus $\sigma$ relation at
intermediate redshift \citep{vandermarel07b}.  Obviously, the spatial
resolution and errors in the kinematic quantities are much larger than
for local galaxies, and consequently, the level of detail is far
lower. Still, the spatially resolved information that is obtained by
going beyond studies of global quantities alone, combined with the
addition of cosmic time to the equation, has already revealed several
interesting results.

\citet{moran07b} used rotation curves to address the question whether
E and S0 galaxies are coeval or not. They found that the relative
number of rotationally supported early-type galaxies at $z\sim 0.5$ is
lower than today, especially in clusters. Combined with extensive
studies of visually classified samples of galaxies out to $z\sim 1$
\citep{dressler97,postman05,smith05}, this is the best evidence to
date for relatively recent S0 formation, presumably through the
transformation of in-falling, star-forming Sa-like galaxies into
quiescent cluster S0 galaxies \citep[e.g.,][]{gunn72, larson80}.

This classical picture does not address the high fraction of S0
galaxies at low densities. Approximately 60\% of all early-type
galaxies in the field are S0 galaxies. Moreover, recently it was shown
that the total fraction of early-type galaxies does not significantly
change between $z\sim 0.8$ and the present either in clusters
\citep{holden07} or in the field \citep{vanderwel07b} if only galaxies
with masses $M\gtrsim 0.5M^*$ are considered. This calls into question
the claim that S0 galaxies evolve differently from E galaxies,
although we note that this has so far not been explicitly addressed.
Because of these reasons it is important to not lose sight of
alternative scenarios for the formation of S0 galaxies. Merging that
does not completely destroy the disk is another proposed mechanism
\citep[e.g.,][]{bekki97}. This is more similar to the favored
formation mechanism for E galaxies and would therefore imply that E
and S0 are in fact one class of objects with a large range in
bulge-to-disk ratios. This ratio may be related to the mass ratio and
the dissipation of the mergers \citep[e.g.,][]{naab06b}. In this
scenario E and S0 galaxies are expected to be roughly coeval.

Another interesting observation in these contexts is that
\citet{vandermarel07a} found some evidence that the relative number of
rotationally supported cluster elliptical galaxies decreases slightly
between $z\sim 0.5$ and the present. This points toward a scenario in
which elliptical galaxies gradually lose angular momentum through
interactions and mergers \citep[e.g.,][]{naab06}. This result is not
necessarily inconsistent with the conclusions from \citet{moran07b}.
The latter authors used rotation to trace the number of S0 galaxies,
whereas the sample of \citet{vandermarel07a} contained (almost) no S0
galaxies by construction. Therefore, these studies probed different
aspects of the early-type (E+S0) galaxy population.

Besides constraining rotation, resolved kinematic data also allow for
a measurement of $M/L$. The evolution of $M/L$ with redshift is
typically studied using estimates based on the virial theorem or the
fundamental plane \citep[e.g.,][]{vandokkum07}. This uses only
globally averaged quantities (luminosity, radius, integrated velocity
dispersion), combined with assumptions about the dynamical structure
of the galaxies and the homology of their evolution.  Spatially
resolved kinematics have the advantage that they allow the $M/L$ to be
inferred through detailed dynamical modeling, which removes some of
the assumptions inherent in these studies.  Systematic effects may
affect not only the measured, average evolution of $M/L$ but also the
evolution of the tilt of the fundamental plane, i.e., the dependence
of $M/L$ evolution on galaxy mass \citep[e.g.,][]{vanderwel05,
  treu05b}.  Van der Marel \& van Dokkum (2007b) addressed this for a
sample of clusters elliptical galaxies at $z \sim 0.5$. They found
that bright, nonrotating galaxies are consistent with homologous
evolution, but that the situation may be more complicated for less
luminous, rotating galaxies. However, the small size of their sample
and the limited number of S0 galaxies left room for different
interpretations of this result.

In this paper we study the resolved kinematics of a sample of field
early-type galaxies at $z \sim 1$, using spatially resolved spectra
from the Very Large Telescope (VLT) and imaging from the
\textit{Hubble Space Telescope} (\textit{HST}). This dataset was
presented earlier by \citet{vanderwel05}.  Our analysis methodology is
largely similar to that of van der Marel \& van Dokkum (2007a,
2007b). However, our sample is different from theirs in three
important areas. First, we focus on field galaxies instead of cluster
galaxies.  Second, our sample includes significant numbers of S0 and
Sa galaxies, instead of just E galaxies. And third, we work at $z \sim
1$ instead of $z \sim 0.5$, thus providing a much extended baseline
for studying evolution.  In particular, we quantify the evolution of
the relative number of rotationally supported field early-type
galaxies from $z\sim 1$ to the present, and we examine the difference
between the $M/L$ inferred from spatially resolved and unresolved
kinematic data.

We have organized the paper as follows. In \S\ref{secdata} we
introduce the data. In \S\ref{secmod} we briefly describe the
dynamical modeling. In \S\ref{secrot} we present our results
regarding rotational support. In \S\ref{secml} we present our
results regarding the $M/L$. In \S\ref{secsum} we summarize the
main conclusions. We adopt the following cosmological parameters:
$(\Omega_M,~\Omega_{\Lambda},~h) = (0.3,~0.7,~0.7)$.

\section{DATA}
\label{secdata}

\subsection{Spectroscopy}
\label{secspec}

Very deep VLT/FORS2 spectroscopy of a sample of magnitude-limited,
morphologically selected field early-type galaxies at redshifts
$0.6<z<1.2$ in the Chandra Deep Field-South (CDF-S) and the foreground
of the $z=1.24$ cluster RDCS 1252.9-2927 (CL1252) were presented by
\citet{vanderwel05}. That paper provides a full description of the
data reduction, including the technique used to measure the global
velocity dispersion of each galaxy from its absorption features. In
this paper we examine a subsample of 25 early-type galaxies (10 E, 9
S0, and 6 Sa galaxies) that have such high-quality spectra that
variations of mean velocity $v$ and velocity dispersion $\sigma$ along
the slit can be measured. The median redshift of this subsample is
$z=0.97$. Since this is a magnitude-limited subsample it is not biased
against rotating or nonrotating galaxies, and it is representative for
the entire population of E+S0 galaxies at this redshift.

\begin{figure}[t]
\epsscale{1.2}
\plotone{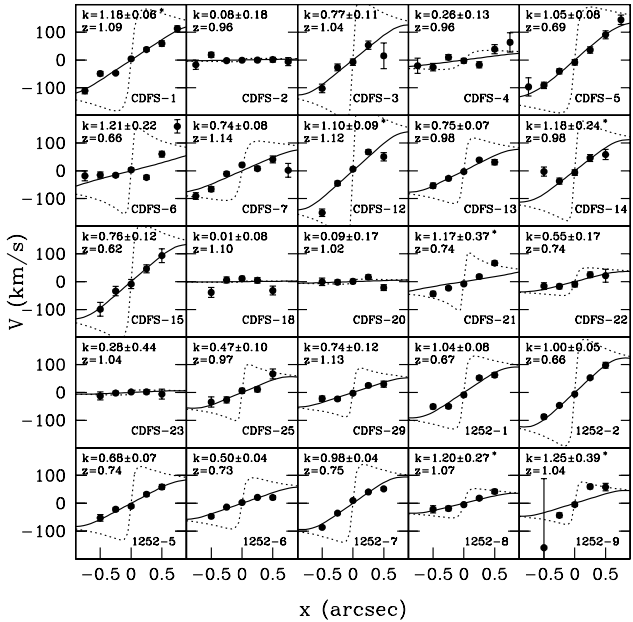}
\caption{Profiles of mean velocity $v$ for 25 field early-type
  galaxies at redshifts $0.6<z<1.2$. The points indicate the measured
  velocity at each spatial pixel in the spectroscopic slits. The solid
  curves show the predictions of the best-fitting models. For
  comparison, dotted curves are the corresponding predictions when
  seeing convolution and pixel/slit binning are not taken into
  account. The visual morphology, the redshift, and the inferred
  rotation parameter $k$ are given for each object. An asterisk
  indicates that the maximum value for $k$ that is physically possible
  is adopted instead of the best-fitting value for $k$. The ID numbers
  correspond to those given in \citet{vanderwel05}, where more
  information regarding the photometric and kinematic properties can
  be found.}
\label{V}
\end{figure}

\begin{figure}[t]
\epsscale{1.2}
\plotone{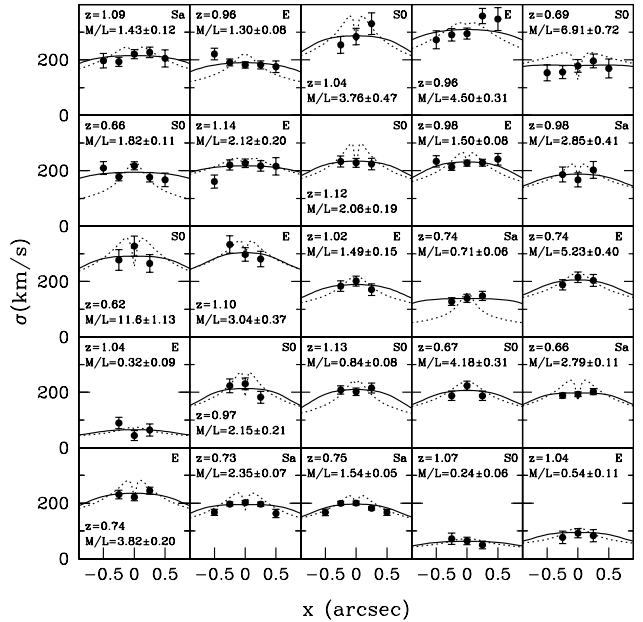}
\caption{Profiles of velocity dispersion $\sigma$ for the same 25
  field early-type galaxies at redshifts $0.6<z<1.2$ as in
  Fig.~\ref{V}. The points indicate the measured dispersion at each
  spatial pixel in the spectroscopic slits. The solid curves show the
  predictions of the best-fitting models.  For comparison, dotted
  curves are the corresponding predictions when seeing convolution and
  pixel/slit binning are not taken into account. The visual
  morphology, the redshift, and the rest-frame $B$-band model $M/L$ in
  solar units are given for each object.}
\label{sig}
\end{figure}

For high-redshift galaxies, $\sigma$ is usually determined from an
extracted spectrum that is averaged over multiple pixels in the
spatial direction in order to optimize for signal-to-noise ratio
($S/N$). Here, instead, we measure $v$ and $\sigma$ for each
individual row of pixels (0.25\" long) for which the $S/N$ is
sufficiently high for a robust measurement. We use the same wavelength
range and template stars as used by \citet{vanderwel05} such that
differences between the results from averaged and resolved spectra are
solely due to differences in the observed spectra themselves.  The
mean velocity profiles, typically measured for 5--7 pixel rows, are
shown in Figure~\ref{V}. It is readily apparent that many galaxies
show signs of significant rotation. Measuring $\sigma$ requires better
$S/N$, therefore $\sigma$ can usually only be measured for 3--5 pixel
rows. The $S/N$ is typically 15 $\rm{\AA}^{-1}$ for the outermost
pixel rows for which we measure $\sigma$ (the central pixel rows
obviously have the highest $S/N$).  For two galaxies (CDFS-18 and
CDFS-23) we have $S/N<8~\rm{\AA}^{-1}$ only for the outermost pixel
rows, which is not ideal for velocity dispersion
measurements. However, for consistency we retained these rows in our
analysis because the large resulting error bars are explicitly
accounted for in our modeling described in \S\ref{secmod}.  In
Figure~\ref{sig} we show the $\sigma$ profiles. In both
Figures~\ref{V} and~\ref{sig} we over-plot the best-fitting models
that we describe in \S\ref{secmod} below.

\subsection{Photometry}
\label{secphot}

GOODS\footnote{http://www.stsci.edu/science/goods/} provides deep,
publicly available \textit{HST}/ACS imaging of the CDF-S
\citep{giavalisco04}.  ACS imaging for the field of CL1252 is also
available \citep{blakeslee03}.  We used these data to assign
morphological classifications (E, S0, Sa) based on visual inspection
of the images, following the strategy outlined by \citet{postman05}.
These classifications are used throughout this paper.

Van der Wel et al. (2005) used the F850LP images to fit de Vaucouleurs
profiles to the objects in their sample in order to determine their
effective radii $R_{\rm{eff}}$ and surface brightnesses
$\mu_{\rm{eff}}$.  For the present analysis we use the ACS data to
measure the full surface brightness profile for each galaxy. We follow
the same procedure described by \citet{vandermarel07a}. First, the
images are deconvolved with the point-spread function (PSF; for which
we use stars in the field) using the Lucy-Robertson
algorithm. Subsequently, we fitted elliptical isophotes to the
two-dimensional images with the IRAF task ELLIPSE. This yields
one-dimensional profiles of major-axis surface brightness, position
angle, and ellipticity. For galaxies at redshifts $z>0.85$ we use the
ACS F850LP images, and for galaxies at redshifts $z<0.85$ we use the
ACS F775W images. With these choices, the observed wavelength
corresponds as closely as possible to the rest-frame $B$ band. The
observed surface brightness profiles are transformed into the
rest-frame $B$ band using the method presented in
\citet{vandokkum07}. This facilitates the comparison with local galaxy
samples.
 
\section{MODELS}
\label{secmod}

For a full description of the modeling procedure we refer to
\citet{vandermarel07a}. In short, the rest-frame $B$-band photometry
described in \S\ref{secphot} is fitted with the projection of a
parameterized, oblate axisymmetric, constant axial-ratio luminosity
distribution. The results of the modeling are fairly insensitive to
variations in the unknown inclination (van der Marel \& van Dokkum
2007a, 2007b). In the discussion below we adopt for each galaxy the
inclination angle that is most likely, given the probability
distribution of intrinsic axial ratios derived from large galaxy
catalogs of the local universe.

Given the three-dimensional luminosity density, the Jeans equations
are solved under the assumptions of a constant $M/L$ and a
two-integral distribution function $f=f(E,L_z)$, where $E$ is the
energy and $L_z$ the angular momentum around the symmetry axis. The
models have $\overline{v_R^2} \equiv \overline{v_z^2}$, so their
velocity distribution is isotropic in the meridional plane. The ratio
of $\overline{v_\phi^2}$ to $\overline{v_R^2}$ is determined by the
requirement of hydrostatic equilibrium. The second azimuthal velocity
moment is split into mean and random components according to the
convenient parameterization
\begin{equation}
\label{satohk}
\overline{v_{\phi}} = k \left(\overline{v_{\phi}^2} - \overline{v_R^2}\right)^{1/2} .
\end{equation}
A value $k=0$ yields a galaxy that is nonrotating and fully pressure
supported. A value $|k|=1$ yields an oblate isotropic rotator, i.e., a
galaxy with sufficient rotation to fully explain its flattening.  The
value of $|k|$ can be larger than unity, with a physical maximum that
depends on the shape of the galaxy, and set by the requirement that
$\sigma_\phi^2 \equiv \overline{v_{\phi}^2} - (\overline{v_{\phi}})^2$
is everywhere positive. In the following we always choose the major
axis position angle (which is indeterminate module $180^{\circ}$) so
that $k$ is positive.

The solutions of the Jeans equations are integrated along the line of
sight, and then convolved in luminosity-weighted sense with the seeing
(typically 0.6\"-0.8\") and with the size of the pixels (0.25\" long)
in the slit (1.0\" wide) through which the galaxies' spectra were
taken, taking the angle between the orientation of the slit and the
major axis into account.  The match of the predictions to the
kinematic data from \S\ref{secspec} is optimized by finding the
best-fitting value of the two available model parameters. The
mass-to-light ratio $M/L$ determines the total amount of rms motion
($\sqrt{v^2+\sigma^2}$) whereas the parameter $k$ determines the
amount of rotational support (which is related to $v/\sigma$). Solid
curves in Figures~\ref{V} and~\ref{sig} show the best model fits. For
comparison, dotted curves show the model predictions if no account is
taken of seeing and pixel/slit binning. Clearly, the latter makes a
considerable difference and it is therefore important that this is
properly modeled. Best-fit values of $k$ are listed in the individual
panels of Figure~\ref{V}.

\begin{figure}
\epsscale{1.2}[t] 
\plotone{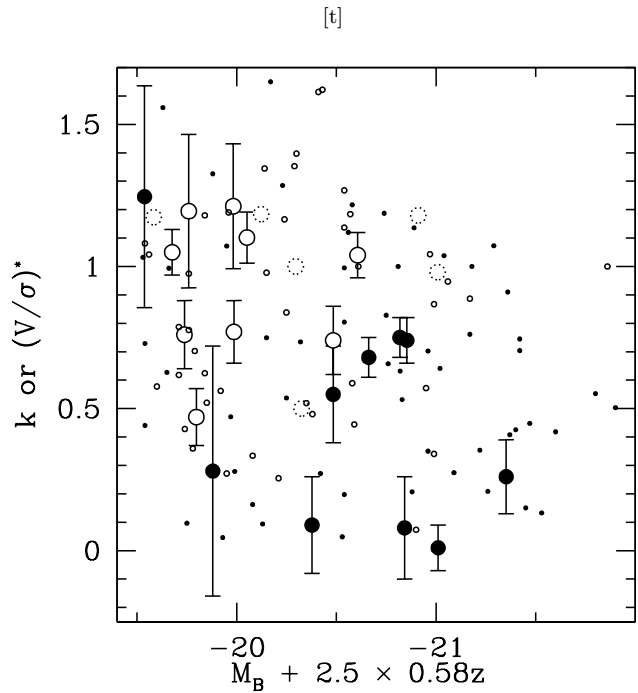}
\caption {$B$-band luminosity vs.~rotation rate. The large symbols are
  the 25 field early-type galaxies at redshifts $0.6<z<1.2$, for which
  we show the rotation parameter $k$. The luminosities are corrected
  for 0.58 dex of evolution between $z=1$ and the present
  \citep{vandokkum07}.  Filled symbols are E galaxies, open circles
  are S0 galaxies, and open, dotted circles without error bars are Sa
  galaxies, as determined by visual classification. The local
  comparison sample is shown as small symbols, with the different
  morphologies distinguished by the same symbols as used for the
  distant sample. $(V/\sigma)^*$, which is comparable to $k$, is used
  to quantify the rotation rate for the local galaxy sample. The
  distributions of the local and distant samples are similar.
\label{L_k}}
\end{figure}

\begin{figure}
\epsscale{1.2}[t] 
\plotone{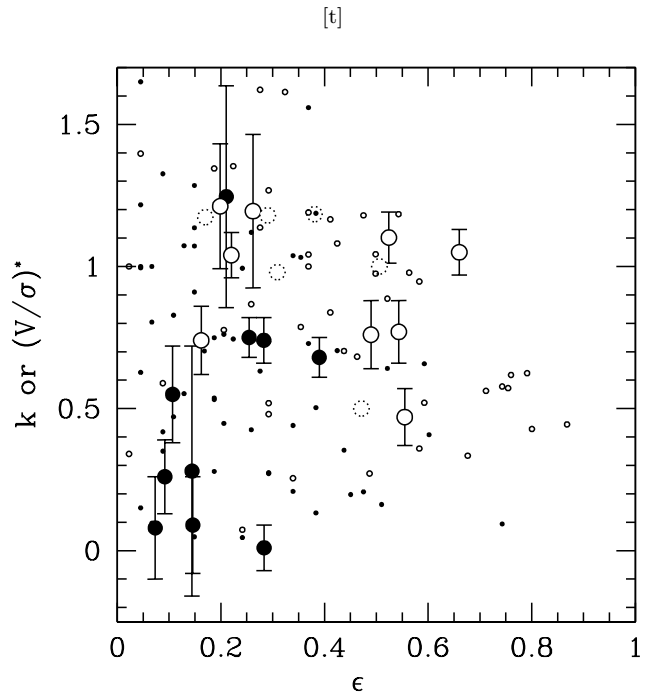}
\caption {Ellipticity vs.~rotation rate [$k$ for the distant sample,
  $(V/\sigma)^*$ for the local sample]. The symbols that indicate
  different visual classifications are the same as in Fig.~\ref{L_k}.
  The ellipticity distribution, like the rotation rate distribution,
  is similar for the local and distant samples.
  \label{e_k}}
\end{figure}

\section{ROTATION AT $z\sim 1$}
\label{secrot}

\subsection{Evolution in Rotation Rate}
\label{rotevol}

As is readily apparent from Figure~\ref{V}, many early-type galaxies
at $z\sim 1$ show signs of rotation. In Figure~\ref{L_k} we show using
large symbols the rotation parameter $k$ as a function of absolute
$B$-band magnitude, distinguishing between galaxies with different
visual morphologies. The absolute magnitudes are corrected for
luminosity evolution using the results of
\citet{vandokkum07}. Slightly different values would have been
reasonable as well (see, e.g., \S\ref{MLimplic}), but none of our
results depend sensitively on this.

Only five out of the 25 galaxies show little or no rotation. These are
all elliptical galaxies. By contrast, all S0 and Sa galaxies show
significant rotation, with many consistent with being rotationally
supported. Figure~\ref{e_k} shows the rotation parameter $k$ as a
function of the apparent ellipticity $\epsilon \equiv 1-(b/a)$, using
similar symbols as in Figure~\ref{L_k}. This shows that the
nonrotating galaxies tend to be rounder than most of the rotating
galaxies.

As noted by \citet{moran07b}, a single-component, spheroidal model may
not be representative for S0 galaxies, some of which clearly show
disks. Our models can in principle be extended to include disks
\citep{cinzano94}, but we have not explored that here. Therefore, the
inferred $k$ may not always necessarily correspond to the physical
property it represents in the model. Nonetheless, as a fit parameter,
$k$ still provides an effective measure of the relative importance of
rotation in the galaxy. In fact, Figures~\ref{L_k} and~\ref{e_k} show
that E and S0 galaxies can be distinguished fairly successfully based
on their kinematics, as quantified by the parameter $k$. Additional
use of luminosity and axial ratio information can further increase the
accuracy of such a kinematical classification scheme, as advocated by
\citet{moran07b}. The only galaxy that seems somewhat out of place in
Figure~\ref{L_k} is the most rapidly rotating elliptical
galaxy. However, this is also the least luminous galaxy in the sample,
and it may well have been visually misclassified.

The clean kinematic separation between different morphological types
at $z\sim 1$ is surprising given the complex situation for local
early-type galaxies \citep{emsellem07}.  Both E and S0 galaxies are
often fast rotators with overlapping kinematic properties
\citep{cappellari07}.  The clean separation at higher redshifts is
therefore quite likely partially artificial.  Morphological
classifications are difficult at high redshift, and recent work has
shown that the relative number of S0 galaxies at high redshift is
systematically underestimated as evidenced by the lack of S0 galaxies
with $\epsilon<0.3$ (B.~P.~Holden et al, in preparation). In addition,
elongated, rotating elliptical galaxies can be misclassified as S0
galaxies.  Nonetheless, these concerns are not relevant for the main
goal of this paper, the evolution of the rotation of early-type
galaxies as a single class of objects.

\begin{figure}[t]
\epsscale{1.2} 
\plotone{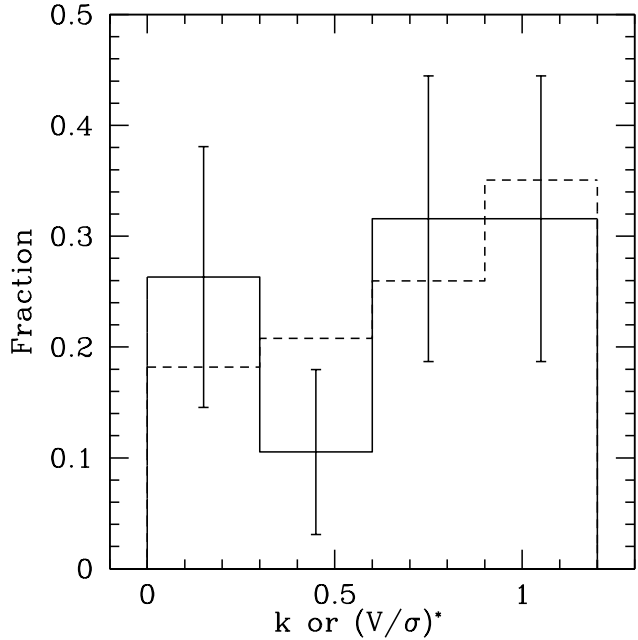}
\caption{Histograms of early-type (E+S0) galaxy rotation rates. The
  parameter $k$ is used for the distant sample (solid line with
  Poisson error bars) and the quantity $(V/\sigma)^*$ is used for the
  local sample (dashed line). Only galaxies brighter than
  $M_{B}=-19.5$ are included (corrected for luminosity evolution,
  in the case of the distant sample). There is no significant
  difference between the two samples in terms of relative numbers of
  slow-rotating and fast-rotating galaxies.}
\label{khist}
\end{figure}

To address the issue of evolution in the rotation rate, we compare the
$z\sim 1$ sample with a local sample extracted from the Lyon
Extragalactic DataBase \citep[LEDA;][]{paturel97}, which is
distributed and made available as HyperLeda
\citep{paturel03}\footnote{http://leda.univ-lyon1.fr}.  We restrict
the local sample to galaxies within $\sim 40$ Mpc (distance modulus
$m-M < 33$). In order to construct a field/group E+S0 sample we
included only galaxies with fewer than 80 neighbors brighter than
$M_{B}=-19.5$ within a cylinder with a 5 Mpc diameter (at the distance
of the galaxy) and with a difference in radial velocity with respect
to the cosmic microwave background of less than $1000\kms$. This
effectively selects a complete sample of 179 galaxies (70 Es, 109 S0s)
outside overdense structures (most notably, the Virgo Cluster).  The
median projected surface density is $\Sigma = 0.4~\rm{Mpc}^{-2}$,
which is derived from the distance to the 7th nearest neighbor down to
the luminosity limit of the magnitude-limited sample with the same
radial velocity within $1000~\kms$.  This is the same value as for a
large, volume-limited sample of galaxies at $z\sim 0.03$
\citep[see][]{vanderwel07b}. It is also comparable with the estimated
surface density (in co-moving units) for a sample of distant field
galaxies \citep[see][]{vanderwel07b} of which our distant sample is a
sub-sample.  Maximum stellar velocities $V$ and velocity dispersions
$\sigma$ are known for 55 Es and 43 S0s \citep{prugniel96}.  There are
37 galaxies in this sample for which multiple, independent
measurements of the rotation velocity $V$ are available.  For 31 of
these the measurements are consistent within the errors.  The
measurements are inconsistent only for the remaining six \citep[see
Table A2 of][]{prugniel96}.  Because especially galaxies with peculiar
properties tend to be targeted multiple times, the measurements for
the remainder of the sample may actually be even better.  Therefore,
the kinematic measurements for the local sample are generally reliable
and reproduceable.

We quantify the rotation rate of the local sample using the quantity
$(V/\sigma)^* = (V/\sigma) \sqrt{(1-\epsilon) / \epsilon}$.  This
approximates the ratio of the actual rotation rate of the galaxy to
the ratio expected for an oblate isotropic rotator of the given
observed axial ratio \citep{davies83}.  This quantity is therefore
similar to our model parameter $k$. The only difference is that $k$ is
defined locally in the meridional plane by equation~(1), whereas
$(V/\sigma)^*$ is defined in terms of the globally defined projected
quantities $V$, $\sigma$ and $\epsilon$. Van der Marel (1991; see his
Fig.~3\textit{a}) made a direct comparison of $k$ and $(V/\sigma)^*$
for a sample of local galaxies and found a good agreement to
$\sim$10\%.  For the present paper we therefore compare $(V/\sigma)^*$
for the local sample directly to $k$ for the distant sample.

The local sample is shown in Figures~\ref{L_k} and~\ref{e_k} using
small symbols. As compared to the distant sample, there are some
differences in the types of galaxies that show high rotation rates. In
the distant sample these are almost all flattened, visually classified
S0 galaxies.  By contrast, in the local sample these include quite a
few roundish, visually classified E galaxies.  It has been known for
some time \citep[e.g.,][]{rixwhite90} that local galaxy catalogs are
deficient in such galaxies. Of course, distant samples may be plagued
by similar, or worse, problems.  However, our sample size is too small
to make definite statements about this issue.

As noted earlier, possible uncertainties in the relative
classifications of E and S0 galaxies do not affect the assessment of
the rotation rate distribution of early-type (E+S0) galaxies in
general.  Figure~\ref{khist} shows that the rotation rate histograms
are statistically indistinguishable for the local and distant samples:
$63\%\pm11\%$ (12 out of 19) of the E+S0 galaxies in the distant
sample have high rotation rates ($k>0.6$), and $59\%\pm5\%$ (58 out of
98) of those in the local sample have high rotation rates
[$(V/\sigma)^*>0.6$].

The similarity between these two numbers is striking, however, they
may not be directly comparable as the kinematic data for the local
sample are not complete, nor representative for the morphological
composition of the population: an E galaxy is twice as likely to have
kinematic data as an S0 galaxy (see the values above). If we, very
crudely, assume that all S0 galaxies rotate and all E galaxies do not,
then the fraction of rotating E+S0 galaxies would be 69\%.  However,
in reality the relative fractions of rotating/nonrotating E and S0
galaxies are not very different (55\% and 65\%, respectively, see the
small data points in Fig.~\ref{L_k}).  Based on these relative numbers
of rotating galaxies for E and S0 galaxies as separate classes we
estimate that the true relative fraction of rotating early-type
galaxies is 61\%, only slightly higher than the measured value of
59\%.

Concluding, we find no evidence for evolution in the rotation rate of
field early-type galaxies between $z=1$ and the present down to the
luminosity limit of the distant sample, $M_{B}=-19.5$.  The distant
sample is small (as evidenced by the large error bars in
Fig.~\ref{khist}), but we would have detected an increase or decrease
by more than 25\% in the relative number of rotating galaxies if there
were such evolution.

\subsection{Comparison with Previous Results}
\label{comprevious}

The only previous measurement of the evolution of the relative number
of rotationally supported early-type (E+S0) galaxies is that of
\citet{moran07b}. Those authors found substantial evolution for
cluster early-type galaxies in the sense that fewer early-type
galaxies are rotationally supported at $z \sim 0.5$ than locally,
which is interpreted as evidence for a lower S0 fraction in the
distant clusters.  This result for cluster galaxies is not directly
comparable to ours. However, the \textit{field} early-type galaxy
sample from \citet{moran07b} contains only $43\% \pm 14\%$ rotating
galaxies, which is less than the value for local galaxies inferred
from the LEDA sample described above.  If this (marginally
significant) decrease between $z\sim 0.5$ and the present is real,
more evolution between $z\sim 1$ and the present can be expected, but
in our sample we do not observe this. Besides the low significance,
this comparison is furthermore hampered by differences in
methodology. Therefore it is interesting to compare our results and
methods with those from \citet{moran07b} in more detail.

Both we and \citet{moran07b} measured spatially resolved kinematical
profiles of galaxies, but beyond that our methods are very different.
\citet{moran07b} infer global quantities (the rotation $v$, velocity
dispersion $\sigma$, and ellipticity), and they do so directly from
the data. Observational effects are not taken into account, and the
internal dynamics of the galaxies are not modeled.  Our method aims to
model the internal dynamics, to take into account known differences
between galaxies (e.g., in their surface brightness profile), and to
account explicitly for all known observational influences on the
measured quantities. The approach of \citet{moran07b} has the
advantage of being simple and model-independent. However, ignoring the
effects of seeing convolution and pixel/slit binning can bias the
results. Figure~\ref{V} shows that these effects have a significant
impact on both the measured rotation value at the outermost radius and
the maximum value of the rotation curve. As a result, the relative
number of rotating galaxies can be underestimated.

To quantitatively illustrate the effect of seeing convolution and
pixel/slit binning, we apply one of the kinematic classifiers of
\citet{moran07b}, namely $v/(1-\epsilon)$, to our datasets. Here, $v$
is defined as half the velocity range of a fitted straight line from
end to end of the measured velocity profile, approximating the maximum
rotation velocity. From the LEDA sample we derive that $65\%\pm5\%$ of
the local population satisfies the criterion from \citet{moran07b},
virtually the same fraction as galaxies with $(V/\sigma)^*>0.6$ (see
\S\ref{rotevol}).  This demonstrates that for the purpose of
distinguishing rotating and nonrotating galaxies, the two methods are
in principle comparable.

In our $z\sim 1$ sample only 7 out of 19 E+S0 galaxies ($37\%\pm11\%$)
have $v/(1-\epsilon)>85 \kms$, consistent with the results from
\citet{moran07b} for $z\sim 0.5$ field galaxies, but substantially
less than the number of galaxies with $k>0.6$ ($63\%\pm11\%$) in our
$z\sim 1$ field sample. However, if we use the maximum model
line-of-sight velocity, with observational effects taken into account
(i.e., the dotted curves instead of the solid curves in Fig.~\ref{V},
then we find a higher fraction with $v/(1-\epsilon)>85\kms$: $63\%\pm
11\%$, the same as the fraction of galaxies with $k>0.6$.  This
implies that omission of the effects of seeing convolution and
pixel/slit binning can lead to significant underestimates of the
number of rotating galaxies at high redshift. This may well explain
the relatively low number of kinematically classified field S0
galaxies found by \citet{moran07b} at $z\sim 0.5$.  However, we cannot
assess the question as to what extent the above-described bias
contributes to the evolution that \citet{moran07b} found for cluster
galaxies.


Interestingly, the kinematic classifications by \citet{moran07b}
agree, statistically speaking, with visual morphological
classifications, at both low and high redshifts.  This argues either
against a bias in their results or that visual classifications can
suffer from a similar bias against rotating/S0 galaxies.  An
indication that high-$z$ classifications of S0 galaxies are indeed
hampered by systematic problems is that the ellipticity distribution
for the cluster early-type galaxy population does not change
significantly with redshift, suggesting that high-$z$ S0 galaxies are
systematically misclassified as E galaxies, more so than the other way
around (B.~P.~Holden et al., in preparation). Still, to what extent biases
contribute to an apparent decline in the S0 fraction with redshift
will probably continue to be a matter of debate for some time to
come. Also, the answer may well be different for cluster and field
galaxies.  However, it is certainly intriguing that our detailed
observations and modeling of field galaxies at $z \sim 1$ show no
evolution in the fraction of rotating early-type galaxies.

Note that our result does not necessarily imply that there is no
evolution in the early-type galaxy population.  On the one hand,
rotationally supported early-type galaxies may merge and produce
pressure-supported elliptical galaxies \citep[e.g.,][]{naab06}. This
process would produce an increase in the fraction of
elongated/rotating galaxies with redshift, for which some tentative
evidence exists in clusters from the results of
\citet{vandermarel07a}. On the other hand, rotating, quiescent
galaxies may be formed out of Sa galaxies that cease to form stars,
contributing to the increase in the stellar mass density of red
galaxies \citep[e.g.,][]{bell04b, brown07} and the decrease of the
cosmic average star formation rate
\citep[e.g.,][]{lefloch05,noeske07a}. As long as the mechanisms that
increase and decrease the relative numbers of rotating and nonrotating
galaxies do not change their ratio, continued growth of both the E and
the S0 population is possible up to the present day.

Neither the results of our own study nor that of \citet{moran07b} are
directly comparable to those obtained by \citet{vandermarel07a} using
the same technique as was in this paper. This is because their study
at $z \sim 0.5$ dealt with a cluster sample that was preselected to
contain almost no S0 galaxies. Therefore, it can shed little light on
the evolution of the S0 fraction with redshift.

\section{MASS-TO-LIGHT RATIOS}\label{secml}
\subsection{Comparison between Model and Virial $M/L$ Estimates}

According to the virial theorem, the mass of a stellar system can be
written as
\begin{equation} 
  M_{\rm{vir}} = \frac{\beta R_{\rm eff} \sigma_{\rm eff}^2}{G} ,
\label{virial}
\end{equation} 
where $R_{\rm eff}$ is the effective radius and $\sigma_{\rm eff}$ is
the luminosity-weighted velocity dispersion measured through an
aperture of size $R_{\rm eff}$. The homology parameter $\beta$
generally depends on the detailed density structure and velocity
dispersion anisotropy of the galaxy. Plausible models can span a wide
range of $\beta$ values. However, \citet{cappellari06} found that
early-type galaxies in the local universe all follow $\beta = 5.0 \pm
0.1$, with surprisingly low scatter. This calibration was obtained by
calculating galaxy masses $M$ from detailed dynamical models for
integral-field kinematical data, including higher-order velocity
moments. The quantity $\sigma_{\rm eff}$ was measured by integration
over the spatially resolved two-dimensional velocity field. The virial
mass-to-light ratio $M/L_{\rm vir}$ can be calculated upon division by
the luminosity $L$. The latter can be measured through aperture
photometry or it can be estimated as $L = 2 \pi R_{\rm eff}^2 I_{\rm
  eff}$, where $I_{\rm eff}$ is the average intensity inside $R_{\rm
  eff}$.

Equation~(\ref{virial}) is commonly used to estimate the mass of
distant galaxies. In doing so, one generally measures the velocity
dispersion $\sigma_c$ through a spectroscopic aperture. The velocity
dispersion $\sigma_{\rm eff}$ is then estimated by applying a
correction formula that is based on typical observations of local
galaxies \citep{jorgensen95}. However, the findings of
\citet{cappellari06} do not guarantee that equation~(\ref{virial})
with $\beta = 5$ is as accurate as it is in the local universe, for at
least two different reasons. First, galaxies may evolve so that the
homology parameter $\beta$ could be a function of redshift. Second,
the accuracy and applicability of the corrections from $\sigma_c$ to
$\sigma_{\rm eff}$ are not guaranteed at high redshift. The
spectroscopic apertures and the seeing in use there often exceed the
galaxy size $R_{\rm eff}$. Observations that adequately mimic this are
not available in the local universe. Moreover, even in the local
universe $\sigma_{\rm eff}/\sigma_c$ shows strong variations from
galaxy to galaxy \citep{cappellari06}. It is therefore necessary to
calibrate equation~(\ref{virial}), and the appropriate value of
$\beta$, at high redshift in the same way as was done locally. This is
possible for our sample by comparing the $M/L_{\rm{Jeans}}$ from our
dynamical models (\S\ref{secmod}) with the values of $M/L_{\rm
  vir}$. Our data quality is obviously not comparable to the
two-dimensional velocity fields and higher-order moments that are
available locally. Nonetheless, this calibration is unique and has not
previously been done at these redshifts.

\begin{figure}[t]
\epsscale{1.2}
\plotone{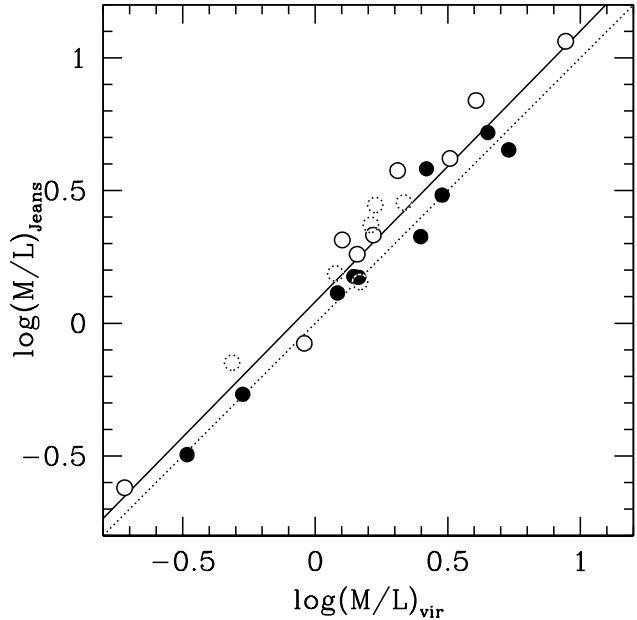}
\caption {Jeans $M/L$ vs.~virial $M/L$ (calculated using $\beta = 5$
  in eq.~[\ref{virial}]) for the $z\sim 1$ galaxy sample. The symbols
  that indicate different visual classifications are the same as in
  Fig.~\ref{L_k}. E galaxies follow the relation $M/L_{\rm Jeans}
  \approx M/L_{\rm vir}$ (dotted line). However, S0 galaxies on
  average have $M/L_{\rm{Jeans}}$ systematically higher than
  $M/L_{\rm{vir}}$ by $\sim 40\%$. The solid line is the least-squares
  fit for the full sample, which has $M/L_{\rm{Jeans}} \propto
  M/L_{\rm vir}^{1.02}$ (and has $M/L_{\rm{Jeans}}$ systematically
  higher than $M/L_{\rm{vir}}$ by $\sim 20\%$).
  \label{ML_ML}}
\end{figure}

\begin{figure*}[t]
\epsscale{1.2} 
\plotone{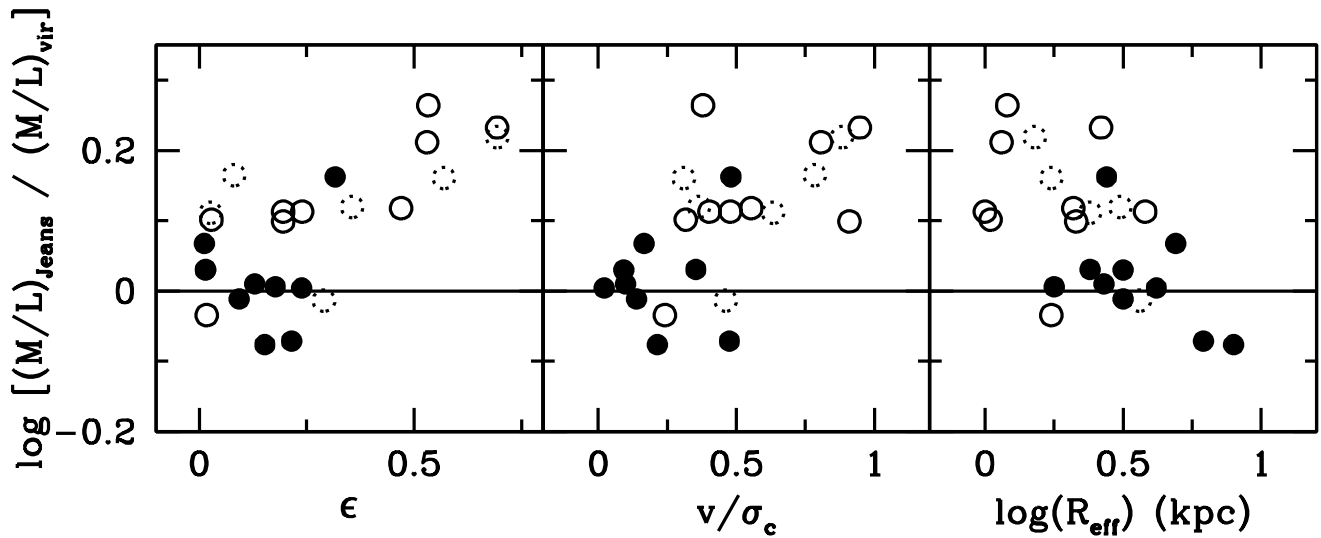}
\caption {The logarithm of the ratio $(M/L)_{\rm{Jeans}} /
  (M/L)_{\rm{vir}}$ vs.~ellipticity (\textit{left panel}), $v/\sigma$
  with $v$ as defined by \citet{moran07b} (\textit{middle panel}), and
  effective radius (\textit{right panel}). The symbols that indicate
  different visual classifications are the same as in
  Fig.~\ref{L_k}. The residuals tend to be positive for galaxies of
  S0 or Sa type, for galaxies with significant ellipticity or rotation
  rate, or for galaxies of small size. Possible causes for this, and
  implications for studies of fundamental plane evolution, are
  discussed in the text.
  \label{var_MLML}}
\end{figure*}

In Figure~\ref{ML_ML} we compare $M/L_{\rm{vir}}$ and
$M/L_{\rm{Jeans}}$ for our sample at $z \sim 1$. We computed
$M/L_{\rm{vir}}$ using $\beta = 5$, with $\sigma_c$ and $R_{\rm{eff}}$
as measured by \citet{vanderwel05}. The spectroscopic aperture was
defined by the width of the spectroscopic slit ($1"$) and the height
of the extracted spectrum (usually, $1.25"$). The ratio
$\sigma_{\rm{eff}}/\sigma_c$ was estimated using the formula of
\citet{cappellari06}, which is consistent with those of
\citet{jorgensen95}. As for local galaxies, the slope of the
best-fitting linear relation between $M/L_{\rm{vir}}$ and
$M/L_{\rm{Jeans}}$ is consistent with unity. However, there is a
systematic offset that appears to depends on the galaxy type. This is
seen in the residuals $\log [(M/L)_{\rm{Jeans}} / (M/L)_{\rm{vir}}]$
shown in Figure~\ref{var_MLML}. For E galaxies, the residuals are
consistent with zero. Therefore, $\beta = 5$ is appropriate for E
galaxies at high redshift, and as a class E galaxies are consistent
with homologous evolution. For S0 (and Sa) galaxies, on the other
hand, $M/L_{\rm{Jeans}}$ is on average almost 40\% higher than
$M/L_{\rm{vir}}$. As shown in Figures~\ref{var_MLML}\textit{a} and
\ref{var_MLML}\textit{b}, these galaxies tend to be flatter and more
rapidly rotating than the E galaxies.  These results are consistent
with the earlier findings of \citet{vandermarel07b} in clusters at $z
\sim 0.5$. However, the trends are much clearer in the present sample
because of its higher fraction of rapidly rotating S0 and Sa galaxies.

There are several possible explanations for the trends observed in
Figure~\ref{var_MLML}. First, S0 and Sa galaxies at $z \sim 1$ may
have $\beta > 5$. However, this would imply that these galaxies evolve
nonhomologously, while E galaxies do not. This seems to us unlikely as
in that case it would be an odd coincidence that $\beta=5$ for all
types of local galaxies.  Second, it is possible that our models have
overestimated the $M/L_{\rm{Jeans}}$ for rapidly rotating
galaxies. Such galaxies may have rapidly rotating disks that are not
well represented by our constant axial ratio oblate models. However,
\citet{cinzano94} addressed this issue explicitly for the well-studied
nearby early-type disk galaxy NGC~2974 and found that a
single-component model overestimated the $M/L$ by only 12\%. This is
not sufficient to explain the trends seen in
Figure~\ref{var_MLML}. Third, it is possible that the virial
equation~(\ref{virial}) for $M/L_{\rm{vir}}$ underestimates the true
mass of rapidly rotating galaxies at high redshift. Rotation
contributes to the hydrostatic support of the galaxy, so if it is
ignored then the $M/L$ will be underestimated. While rotation does
increase $\sigma_{\rm eff}$ through the effect of line broadening,
this may not fully capture the true hydrostatic importance of the
rotation component. This may be especially important if the seeing is
of order of or larger than the scale at which rotation is manifest.
Such problems are avoided for local galaxies for which much higher
quality data are available.  This may explain the apparent differences
between the $M/L$ of rotating galaxies at low and high redshifts.  For
our distant galaxies, correction formulae for $\sigma_{\rm
  eff}/\sigma_c$ are used that are not themselves calibrated at high
redshift.

\subsection{Implications for the Evolution of $M/L$}
\label{MLimplic}

In the previous section we demonstrated that $M/L_{\rm{vir}}$ may be
too low for rapidly rotating galaxies at higher redshifts, but not for
local early-type galaxies. Since the fundamental plane is in essence a
correlation between $M/L_{\rm{vir}}$ and other global galaxy
parameters, this will bias the amount of $M/L$ evolution inferred from
fundamental plane studies \citep[see][and references
therein]{vandokkum07}.  Van der Wel (2005) found that the rate of
luminosity evolution for the sample of field galaxies that we also use
in this paper is $\Delta \log(M/L) = (-0.76 \pm 0.07)
z$. Figures~\ref{ML_ML} and~\ref{var_MLML} suggest that, averaged over
the rotating and nonrotating galaxies in our sample, $M/L_{\rm vir}$
may be too low by $\sim 20$\%. If so, then the evolution in $M/L$ is
reduced to $\log(M/L) = (-0.69 \pm 0.07) z$. This is steeper than the
evolution found for cluster galaxies, but not by much
\citep{vandokkum07}.

The difference between $M/L_{\rm{Jeans}}$ and $M/L_{\rm{vir}}$ depends
on the galaxy rotation rate. Moreover, rotating galaxies tend to have
smaller $R_{\rm{eff}}$ than nonrotating galaxies (see
Fig.~\ref{var_MLML}\textit{c}). Therefore, our results also have
consequences for the measured evolution of the tilt of the fundamental
plane and hence for the slope in the relation between $M/L$ and either
$M$ or $\sigma$ (as discussed previously by van der Marel \& van
Dokkum 2007b). Evolution in the tilt of the fundamental plane has been
observed by many authors \citep{vanderwel04, treu05b, vanderwel05,
  diserego05, jorgensen05}.  However, it is a matter of debate to what
extent sample selection effects contribute to the observed evolution
\citep{vanderwel05, treu05b}. To illustrate the consequences of the
results presented in this paper we revisit the analysis from
\citet{vanderwel05}.  They showed that the probability that the
observed distribution of $M_{\rm{vir}}$ and $M/L_{\rm{vir}}$ is drawn
from a parent population with the same distribution as the local
galaxy population, apart from a constant amount of luminosity
evolution inferred from the most massive galaxies, is as low as
$0.14\%$.  We repeat the analysis but now using the distribution of
$M_{\rm{Jeans}}$ and $M/L_{\rm{Jeans}}$ and find that this increases
the probability to 3.0\%.  In other words, the evidence for
mass-dependent evolution of $M/L$ remains strong, but it does become
weaker.

These arguments illustrate that the often observed evolution of the
tilt of the fundamental plane is not necessarily entirely due to
mass-dependent evolution of the $M/L$, i.e., downsizing. In order to
determine to what extent the slope and the scatter of the relation
between $M/L$ and $M$ evolve, still deeper observations are required
to overcome the biases caused by sample selection effects in the
surveys conducted over the past few years. In addition, it may be
necessary to use dynamical modeling, as we do in this paper, to
overcome the shortcomings of virial and fundamental plane mass
estimates.

\section{CONCLUSIONS}
\label{secsum}

We use the spatial information of our previously published VLT/FORS2
absorption-line spectroscopy to measure mean stellar velocity and
velocity dispersion profiles of 25 field early-type galaxies in the
redshift range $0.6<z<1.2$, with median redshift $z=0.97$. The
kinematical profiles can be reliably measured even in the most distant
galaxies. Rotation is detected in the majority of the sample. Surface
brightness profiles are determined from \textit{HST}
imaging. Two-integral solutions of the Jeans equations for oblate
axisymmetric, constant axial-ratio models are calculated to interpret
the data, taking into account line-of-sight projection, seeing
convolution, and pixel/slit binning. This yields for each galaxy the
degree of rotational support, as quantified by the parameter $k$, and
the mass-to-light ratio $M/L_{\rm{Jeans}}$ \citep{vandermarel07a}.

The rapidly rotating ($k \gta 0.6$) and slow-rotating ($k \lta 0.6$)
galaxies in the distant sample tend to have different global
properties.  The rapidly rotating galaxies tend to have later
morphological types (S0 or Sa) and tend to be less luminous and more
elongated.  The slow-rotating galaxies tend to have earlier
morphological types (E) and tend to be more luminous and rounder. The
systematic variations with luminosity and axial ratio suggest that the
correlation with morphological type is real, in agreement with the
findings of \citet{moran07b}.  Local E and S0 galaxies show a more
complex and overlapping set of kinematic properties
\citep[e.g.,][]{cappellari07}.  The contrasting clean separation of
rotating and nonrotating galaxies according morphological type
observed at $z\sim 1$ is most likely partially artificial and must be
related to the fact that round/face-on S0 galaxies are often
misclassified as E galaxies, at high redshift even more so than at low
redshift.

We have compiled a local comparison sample to study evolution of the
field early-type (E+S0) galaxy rotation rate. The distribution of $k$
in our $z\sim 1$ sample is statistically indistinguishable from the
distribution of $(V/\sigma)^*$, a comparable measure of rotational
support, in the local sample. The relative fraction of rotating
galaxies does not change significantly between $z\sim 1$
($63\%\pm11\%$) and the present ($61\%\pm5\%$). If rotation is taken
to be a reliable indicator of morphological type, then this provides
evidence for an unchanging fraction of S0 galaxies in the early-type
{\it field} galaxy population with redshift. This conflicts with the
findings of \citet{moran07b} who did not correct for the effects of
seeing convolution and pixel/slit binning on the measured rotation
curves. It is possible that this may have led to an underestimate of
the number of rapidly rotating field galaxies at $z\sim 0.5$ in their
study.

We have compared the mass-to-light ratio $M/L_{\rm{Jeans}}$ from our
spatially resolved models to the values $M/L_{\rm{vir}}$ inferred by
applying the virial theorem to globally averaged quantities.  For E
galaxies, which are generally slow-rotating, we find good agreement
using a homology parameter $\beta = 5$, which is consistent with the
value calibrated locally. So there is no evidence for non-homologous
evolution of E galaxies out to $z \sim 1$. On the other hand, for
elongated, rotating galaxies (which are often S0s) we find that
$M/L_{\rm{Jeans}}$ is on average $\sim 40\%$ higher than
$M/L_{\rm{vir}}$ (see Figs.~\ref{ML_ML} and \ref{var_MLML}).  For
local galaxies this trend is not observed. This may hint at
non-homologous evolution of this galaxy population ($\beta$ increasing
with redshift); it may suggest that our dynamical models produce
biased results in these galaxies because of the neglect of cold disks;
or it may suggest that the virial formula for $M/L$ produces biased
results when applied to poorly resolved, rapidly rotating galaxies at
high redshift. We cannot unambiguously identify the true cause, but to
us the latter seems to be the most straightforward explanation. If so,
then studies of fundamental plane evolution overestimate both the
amount of evolution and the evolution in the tilt of the fundamental
plane (and the relation between $M$ and $M/L$), which is generally
interpreted as evidence of down-sizing.

\acknowledgements{The authors would like to thank Pieter van Dokkum,
  Marijn Franx, Michele Cappellari, Dan Kelson, and Stijn Wuyts for
  helpful discussions and suggestions.  A.~v.~d.~W.~acknowledges
  support from NASA grant NAG5-7697.}

\end{document}